\documentclass[a4paper]{jpconf}
\usepackage{graphicx}
\begin{document}
\title{Matter-Antimatter Asymmetry in the Universe via String-Inspired CPT Violation at Early Eras}

\author{Nick E. Mavromatos}

\address{Theoretical Particle Physics and Cosmology Group, Physics Department, King's College London, Strand, London, WC2R~2LS, UK}

\ead{nikolaos.mavromatos@kcl.ac.uk}

\begin{abstract}
In four-space-time dimensional string/brane theory, obtained either through compactification of the extra spatial dimensions, or by appropriate restriction to brane worlds with three large spatial dimensions, the rich physics potential associated with the presence of non-trivial Kalb-Ramond (KR) axion-like  fields has not been fully exploited  so far. In this talk, I discuss a scenario whereby such fields produce spontaneous Lorentz- and CPT-violating cosmological backgrounds over which strings propagate, which in the early Universe can lead to Baryogenesis through Leptogenesis in models with heavy right-handed neutrinos. 
\end{abstract}

\section{Introduction and Motivation}

 Despite its enormous success, the Standard Model (SM) of particle physics fails to provide an explanation as to why we exist, more specifically, why matter dominates over antimatter in the Universe. According to observations of the Cosmic Microwave 
 Background (CMB) radiation~\cite{planck}, the baryon asymmetry is measured to be:
 \begin{equation}\label{bau}
Y_{\Delta B} \equiv \frac{n_{B}-n_{\bar{B}}}{n_{\gamma}}=(6.1\pm0.3)\times10^{-10}
 \end{equation}
where $n_{B}$ is the number density of baryons, $n_{\bar{B}}$ is
the number density of antibaryons and $n_{\gamma}$ is the density of photons.

The baryon asymmetry (\ref{bau})  cannot be explained \emph{quantitatively} within the SM. Indeed, the latter is a theory that predicts that matter and antimatter have been created in equal amounts after the Big-Bang, as a consequence of the CPT theorem~\cite{cpt}. According to it, the discrete symmetry transformations of Charge Conjugation (C), Parity (Spatial Reflexion, P) and Reversal in Time (antiunitary, T), when applied in any order, leave the action (and hence all dynamics derived from it) invariant, provided that the corresponding quantum field theory is Lorentz invariant, has interactions that respect locality and is unitary, that is, it has a Hermitian Hamiltonian (and the associated scattering matrix is well defined). 
To explain the observed matter antimatter asymmetry in the Universe in a way consistent with the above theorem, Sakharov~\cite{sakharov} postulated that any local CPT invariant quantum field theory that was to describe 
 the early Universe and the observed baryon asymmetry must have interactions that \emph{violate} Baryon Number (B), Charge Conjugation Symmetry (C) and Charge-Parity (CP) symmetry. There should also be a departure from thermal equilibrium, since in a CPT invariant theory in thermal equilibrium any Baryon number violation will be washed out as a consequence of the Boltzmann exponential factor. Although the SM contains qualitatively all such features, for instance, Baryon number is violated due to quantum anomalies, and CP Violation (CPV) appears as a complex phase in the quark mixing matrix of the weak interactions\footnote{CP violation was observed in fact for the first time in the meson sector in neutral Kaon experiments~\cite{cpviol}.There may also be CP violation in the lepton sector (neutrino mixing matrix) but this is still unobserved.}, nevertheless the generated matter-antimatter asymmetry in the SM is smaller than the observed one by several orders of magnitude~\cite{rubakov,smbau}. Hence, to explain the observed dominance of matter over antimatter in the Cosmos we should look for Physics beyond the SM (BSM), such as supersymmetry, extra dimensional models, string theory \emph{etc.}, in an attempt to identify the necessary extra sources of CPV that would produce the asymmetry between baryons and antibaryons.

 In fact, the observed baryon asymmetry can be generated either directly, through CPV out-of-thermal-equilibrium processes in the baryon sector in the early Universe (Baryogenesis)~\cite{baryo}, or first through out-of-thermal-equilibrium CPV processes in the Lepton sector, often associated with CPV decays of heavy right-handed neutrinos (RHN) into SM particles, which produce a Lepton asymmetry (Leptogenesis)~\cite{lepto,lepto2}. The latter is then communicated to the baryon sector through sphaleron processes~\cite{rubakov} that violate both Lepton (L) and Baryon (B) numbers but  conserve the difference B-L. 
A minimal extension of the standard model, with two or three generations of heavy Majorana RHN, with a Higgs portal that allows for CPV decays of the heavy neutrinos into lepton-Higgs pairs in the early Universe, suffices to generate baryogenesis through leptogenesis~\cite{FY,Luty}. It is important to realise that in such scenarios the required CPV in the neutrino sector appears  in the relevant mixing matrix, and as such the model
cannot work with a single generation of RHN. 
Nevertheless, the presence of more than one generation of heavy neutrinos is required by the see-saw mechanism which is responsible for generating masses for the light neutrinos of the SM sector~\cite{seesaw}. The difference in the widths of the decay of the heavy (Majorana) neutrinos into SM leptons as compared with the corresponding decays into antileptons, which is essential for leptogenesis, occurs at one loop level in these models. We should remark at this point, however, that there is a fair amount of fine tuning of parameters in the pertinent models that needs to be done in order for the widths of the appropriate CPV decays to produce phenomenologically realistic values for baryogenesis.

 A fundamental paradigm shift for the generation of baryon asymmetry would be to relax CPT Invariance in the very early Universe. In a simplest setting, several models that include \emph{spontaneous} Violation of Lorentz symmetry (LV) may imply CPT Violation (CPTV)~\cite{greenberg}, provided that the other two requirements of the CPT theorem, locality and unitarity\footnote{The relaxation of any one of these requirements alone would constitute independent reasons for CPTV~\cite{novikov}.}, are obeyed.
Such Lorentz- and CPT-Violating models  can be parameterised in the context  of the so-called Standard Model Extension (SME)~\cite{sme}. Today, there are very stringent bounds on the amount of the allowed LV and CPTV, as inferred by a plethora of diverse precision measurements~\cite{smebounds}.  However, in the early Universe such effects could be much stronger, and in fact it was proposed in \cite{Bertolami} that baryogenesis can occur within the context of string-inspired SME models~\cite{smestring}, in which structures corresponding to tensorial quantities of order-$k$, which acquire non-trivial vacuum expectation values (v.e.v), $\langle T^{(k)} _{\mu_1 \dots \mu_k} \rangle  \ne 0$, provide 
 a sort of chemical potential $\mu$ and energy splitting between particles and antiparticles (quarks-antiquarks specifically). For instance, such string-inspired SME models include interactions of the form $ \mathcal{L}_I \ni \frac{\lambda \, \langle T^{(k)}_{\dots } \rangle}{M_s^k}  \overline \psi (\gamma^0)^{k +1} (i \partial_0 )^k \psi + {\rm h.c} $, where $\psi$ are fermion fields (e.g. quarks), $M_s$ is some high (e.g. string) mass scale and it is expected generically that $\langle T^{(k)} \rangle = {\mathcal O}\Big((m_{\rm low}/M_s)^\ell \, M_s\Big)$, where $\ell$ is some positive power, and $m_{\rm low}$ is a characteristic low-energy scale.
Dominant terms have $k + \ell = 2$.  As argued in \cite{Bertolami}, such interactions can generate  
a difference between the thermal equilibrium densities of quarks and antiquarks\footnote{We note that if CPT is violated, the Sakharov condition of out-of-equilibrium processes can be alleviated.}, e.g. for the case $k=0, \ell=2$: $\frac{n_q - n_{\bar q}}{s} = \frac{45\, g_q}{2\pi^4\, g_s(T)} \frac{\mu}{T}\, I_0(m_q/T)$, with $I_0(x) = \int_x^\infty dy \, y \, \sqrt{y^2 - x^2} \, \frac{e^y}{(1 + e^y)^2} $, $g_q$ the number of internal quark degrees of freedom, 
$s= \frac{2\pi^2}{45}\, g_s{(T)}\, T^3$ the entropy density and 
$g_s(T) = \sum_{\rm B(osons)} g_B (\frac{T_B}{T})^3 +  \frac{7}{8} \, \sum_{F(ermions)}\, g_F\,  (\frac{T_F}{T})^3$ the total entropy degrees of freedom, assuming standard cosmology, with $T_{B(F)}$ the corresponding decoupling temperatures, and $g_{B(F)}$ the degrees of freedom, of Boson (Fermion) species. The induced chemical potential in this case is of order $\mu  \sim m_{\rm low}^2/M_s$. Such considerations have been argued~\cite{Bertolami} to lead to a baryon asymmetry which is strong in the early Universe, but diluted today by means of sphaleron processes. The model is interesting, although, rather generic in the sense that the tensor fields which acquire non trivial v.e.v have not been specified within phenomenologically realistic microscopic string models.

\section{A Lorentz and CPT Violating String-Inspired Model for Leptogenesis}

It is the purpose of this talk to describe a class of detailed, and phenomenologically realistic, string-inspired quantum field theory models which can lead to efficient baryogenesis through leptogenesis within 
conventional string theory frameworks~\cite{dms}. The latter are characterised by \emph{spontaneous} Lorentz and CPT Violation through propagation of strings 
over specific LV and CPTV backgrounds that 
satisfy the world-sheet conformal invariance conditions and hence constitute acceptable vacua of string theory. 
The backgrounds are of cosmological type, that is the corresponding fields depend only on the cosmic time~\cite{aben}. The background field that yields the spontaneous Lorentz and CPT Violation is the spin-one antisymmetric tensor Kalb-Ramond (KR) field $B_{\mu\nu} = - B_{\nu\mu}$ of the bosonic massless gravitational string multiplet~\cite{sloan,sloan2,sloan3}. Its field strength is a totally antisymmetric three-tensor
$H_{\mu\nu\rho} = \partial_\mu B_{\nu\rho} + {\rm (cyclic~permutation~of~the~indices)}$. In four space-time dimensions, this field strength is dual to a pseudoscalar (KR ``axion'') field $b^{\rm KR}(x)$:
$H_{\mu\nu\rho} = e^{2\Phi} \epsilon_{\mu\nu\rho}^{\,\,\,\,\, \,\,\,\,\,\,\sigma}\, \partial_\sigma (b^{\rm KR})$, 
where $\epsilon_{\mu\nu\rho\sigma}$ is the covariant Levi-Civita tensor (in the presence of gravity), and $\Phi$ is the dilaton field (scalar field of the bosonic massless  gravitational multiplet of the string). 
In the absence of other matter fields, there is a consistent cosmological solution~\cite{aben} of a linearly(with the cosmic time $t$)-expanding Robertson-Walker Universe, which is valid to all orders in the string Regge slope $\alpha^\prime = 1/M_s^2$. In the Einstein frame, where the (lowest-order in derivatives) scalar curvature term in the effective action is canonically normalised as in the Einstein -Hilbert action, the solution is characterised by a logarithmically (in $t$) dependent dilaton  and a linear in $t$ KR axion field~\cite{aben}
\begin{equation}\label{bs}
b^{\rm KR} (t) = c_1 \, t 
\end{equation}
where the proportionality constant $c_1$ depends on microscopic data of the underlying string theory. 
To first order in $\alpha^\prime$, that is up to and including second order derivative terms of the various fields, which we restrict our attention to here, the target-space effective action obtained from this string theory is\footnote{From now on we also assume that the dilaton has been stabilised by its (string-loop induced) potential to an appropriate constant value, which by normalisation we set to $\Phi=0$.}~\cite{dms,aben,sloan,sloan2,sloan3}:
\begin{eqnarray}\label{action}
\mathcal S = \int d^4 x \, \sqrt{-g} \Big[\frac{1}{2\kappa^2} R - \frac{1}{2}(\partial_\mu b^{\rm KR})^2 - \Omega 
+ \sum_i [{\overline \psi}_i (i\gamma^\mu \partial_\mu -m_i) \psi_i + \frac{\kappa}{3\sqrt{6}} \partial_\mu b^{\rm KR}\,  {\overline \psi}_i \gamma^\mu  \gamma^5 \psi_i ] + \dots \Big]
\end{eqnarray}
where the sum is over fermion species in the theory, including those in the SM sector, $\kappa = 8\pi M_P^{-2}$ is the gravitational constant in four space-time dimensions, with $M_P$ the Planck mass scale, and $\Omega$ is a vacuum energy arising from (a plethora of terms of) the underlying microscopic string theory. In several string/brane models of the Universe, the vacuum energy $\Omega$ contains negative (anti de Sitter) parts (e.g. coming from bulk contributions in brane world scenarios~\cite{rizos}), which can be used to suppress the positive vacuum energy due to the kinetic terms of the KR axion in the background (\ref{bs}). Such suppression is important in ensuring standard cosmology during the radiation era, a result which is used in \cite{dms} and here.
The coupling of the KR axion field $b^{\rm KR}(x)$ with the fermions is dictated by the interpretation of the field strength $H_{\mu\nu\rho}$ as a totally antisymmetric part of a torsion in a generalised connection, given that the kinetic terms of $b^{\rm KR}(x)$ field and the curvature terms $R$ in the effective action (\ref{action}) can be expressed in terms of a generalised curvature scalar of such a torsionful connection~\cite{sloan,sloan2,sloan3}.  The dots ($\dots $) in (\ref{action}) denote gauge and other terms in the effective theory pertaining to the SM sector, but also higher derivative terms arising from strings, which are ignored for our low energy purposes. 

\begin{figure}
\begin{center}
\includegraphics[width=0.3\columnwidth]{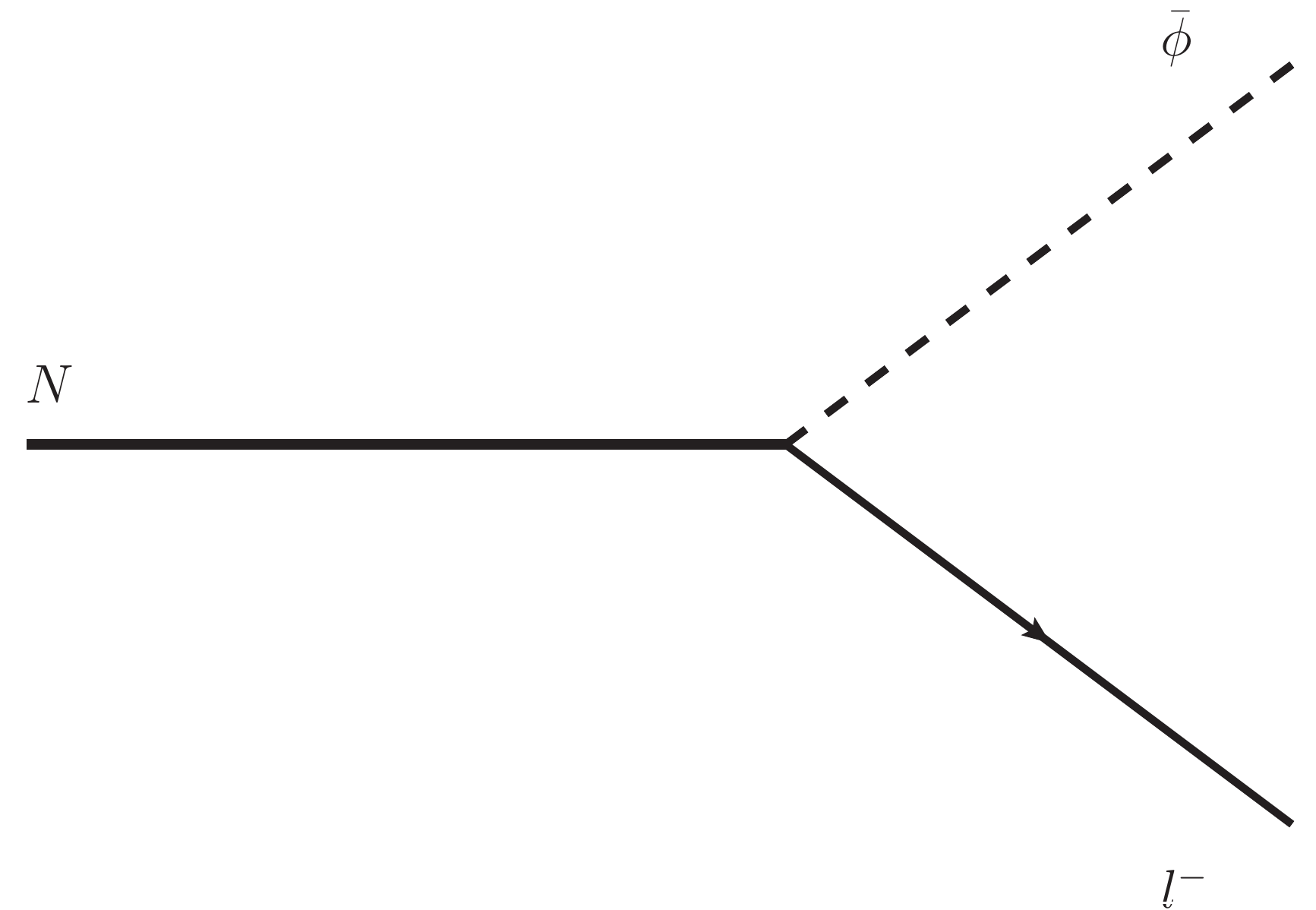} \hspace{2cm}
\includegraphics[width=0.3\columnwidth]{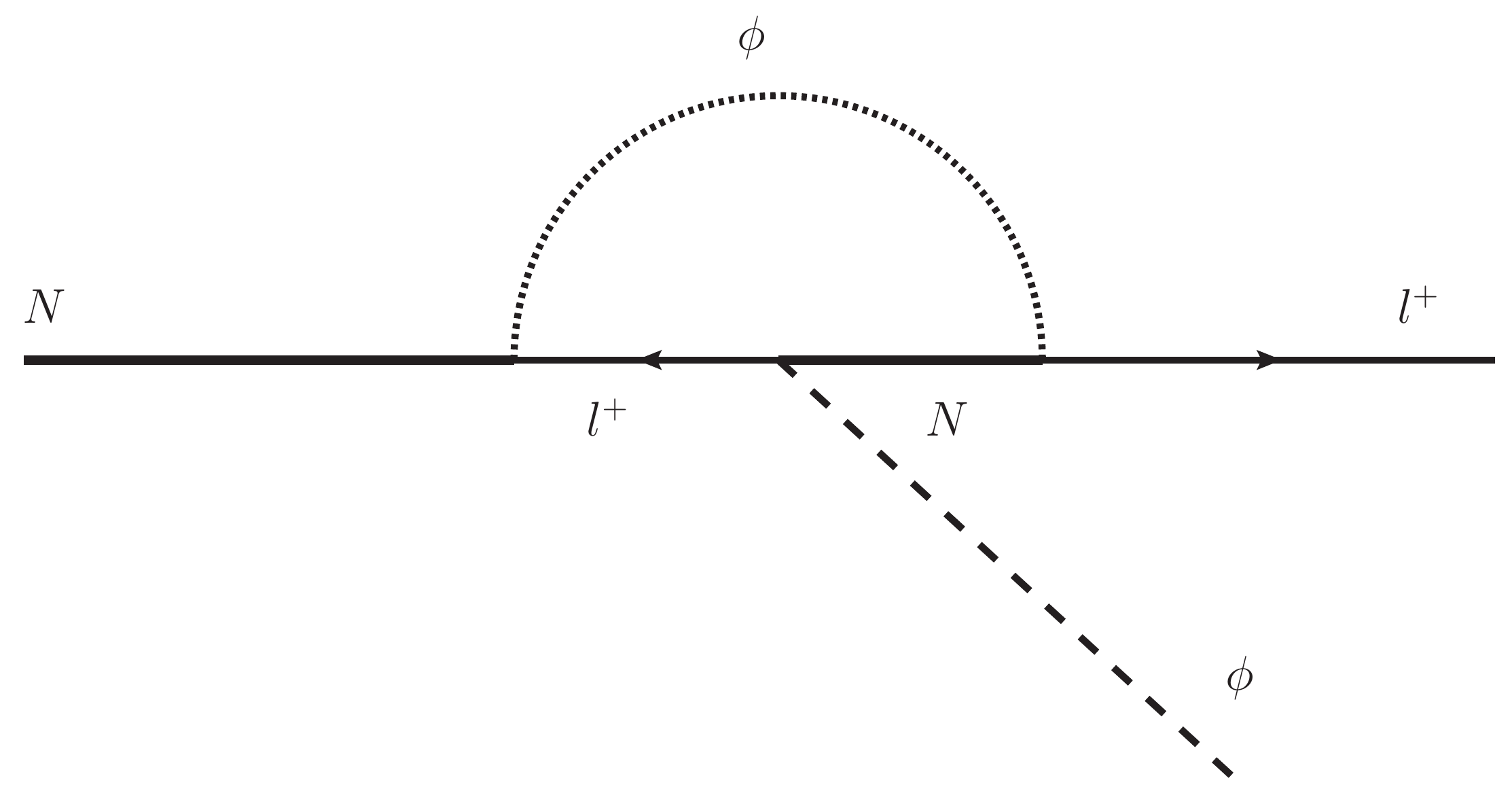} 
\end{center}
\caption{Tree- (left) and one-loop (right) decay amplitudes for the leptogenensis decays (\ref{2channels}). }
\label{fig:decays}
\end{figure}

In the presence of fermionic and other matter, the background solution (\ref{bs}), which implies a constant field strength, $H_{ijk} = {\rm const}, i,j,k=1,2,3,$ is a valid solution provided there is a constant temporal axial vector condensate of fermions that violates Lorentz and CPT symmetry but  respects the isotropy of the Universe~\cite{dms}, \begin{equation}\label{cond}
\sum_i \, \langle  {\overline \psi}_i \gamma^0  \gamma^5 \psi_i \rangle \ne 0~.
\end{equation}  
In case of a constant $\partial_t b^{\rm KR} \equiv {\dot b}^{\rm KR}$ background, the effective action in the fermionic sector 
acquires the form of a SME with a non-trivial LV and CPTV axial vector background over which fermions propagate, $b_\mu \sum_i  \, \overline \psi_i \gamma^\mu \gamma^5 \psi_i $, with $b_\mu \equiv \partial_\mu b^{\rm KR} $ (with only $b_0 \ne 0$~\footnote{In a quantum path integral based on the model (\ref{action}), this can be seen by integrating exactly the $\partial b^{\rm KR}$ -depenendent terms, which yields~\cite{sloan3} repulsive four-fermion interactions $J_\mu^5 \, J^{5\, \mu}$ of the axial current $J_\mu^5 = \sum_i \overline \psi_i \gamma_\mu \gamma^5 \psi_i $. In string effective theories, at finite temperature and density, there may be higher order (attractive) fermion  contact interaction terms arising from the exchange of heavy string states, that may lead to a condensate $\langle J_\mu^5 \rangle$ of the form (\ref{cond}) about which one expands the fermion quantum fluctuations. An SME-like effective action follows then from (\ref{action}), along with (de Sitter) type effective contributions to the vacuum energy. Moreover, in this language, it is seen immediately that, upon its destruction, the condensate (being comprised of ordinary matter) will scale with the temperature as $T^3$~\cite{dms}.}). 
The model of \cite{dms} includes  a single species of massive Majorana RHN, $N_R$, which also couple to the KR axion via the axial vector fermion coupling of (\ref{action}). 
However, for such Majorana neutrino any axial condensate vanishes, in contrast to other chiral fermions of the SM, including left handed active neutrinos, which contribute to (\ref{cond}). 
The RHN part of the effective action in the KR axion background (\ref{bs}) reads:\\
$\qquad \qquad  \frac{\mathcal{L}^N}{\sqrt{-g}} = i\overline{N_R}\gamma^\mu {\partial_\mu}N_R-\frac{M}{2}(\overline{N_R^{c}}N_R+\overline{N_R}N_R^{c}) +  b_\mu \overline{N_R}  \, \gamma^\mu \, \gamma^{5}N_R-y_{1}\overline{L}_{1}\tilde{\phi}N_R+h.c.,$ \\ 
where $N_R$ is the RHN field (with its Majorana field being given by $\nu \equiv N_R + N_R^c$), $M$ is a Majorana mass,  $\tilde \phi$ is the adjoint of the Higgs field,  $\tilde{\phi}_i=\varepsilon_{ij}\phi_j $, and $L_{1}$ is a lepton (doublet) field of the Standard Model (SM) sector in the first generation, with $y_1$ a Yukawa coupling, which, if non zero, provides a non-trivial interaction between the RHN and the SM sector (``Higgs portal''). In the presence of the LV and CPTV background (\ref{bs}), we have (\emph{cf.} (\ref{action})): $b_0 = \frac{\kappa}{3\sqrt{6}}\,  {\dot b}^{\rm KR} = {\rm constant } \ne 0$. The portal yields inequivalent decay rates  for the channels~\cite{dms}: 
\begin{eqnarray}\label{2channels}
{\rm Channel ~I~:} \quad  N_R \rightarrow l^{-}h^{+}~, \qquad 
{\rm Channel ~II~:} \quad  N_R \rightarrow l^{+}h^{-}~,
\end{eqnarray}
where $l^\mp$ ($h^\pm$) denote charged lepton (Higgs) fields. It is important to realise that this difference occurs already at tree level and for a single species of heavy RHN, in contrast to the conventional CPV but CPT conserving case~\cite{FY,Luty}, where the corresponding differences due to CPV occur at one loop (\emph{cf.} Fig. \ref{fig:decays}) and require more than one species of RHN, given that the CPV appears in the phases of the neutrino mixing matrix.

As discussed in detail in \cite{dms}, on assuming  $y_1 \sim 10^{-5}$ for concreteness, as well as standard cosmology during the radiation-dominance era, at which the decoupling  of the heavy neutrinos occurs at a temperature $T_D \simeq M$, one can estimate the lepton asymmetry $\Delta L^{TOT}$ generated at $T_D$ as:
$\frac{\Delta L^{TOT}}{s} (T \simeq T_D \simeq M)= \frac{2\Omega b_0}{\Omega^2+b_0^2} \frac{\bar n_N^{eq}(T_D)}{s (T_D)}$,
where $s $ is the entropy density, defined previously, $\Omega^2 = M^2 + b_0^2 $, and $n_N^{eq} (T) =g_N \, e^{-m_N/T}\,\left(\frac{M \, T}{2\pi }\right)^{\frac{3}{2}}+\mathcal{O}(b_0^2/M^2)$   is the thermal equilibrium density of the right-handed neutrinos, with $g_N$ the pertinent internal degrees of freedom. We assumed that $b_0 \ll T_D, M$. This is a self consistent assumption, given that in models of baryogenesis through leptogenesis~\cite{FY,Luty}, as in our model, the lepton asymmetry is expected to be of the same order as the baryon asymmetry (\ref{bau}), and in our case this is achieved for ~\cite{dms}: 
\begin{equation}\label{values}
\frac{b_0}{M} \simeq 10^{-8}~, \quad {\rm with} \quad M \simeq T_D \sim 10^5~{\rm GeV}, \quad y_1 \sim 10^{-5}.
\end{equation}
In \cite{dms} a scenario was also presented for the cooling law of ${\dot b}^{\rm KR}$ with the temperature, by assuming that the axial fermion condensate (\ref{cond}) is destroyed at temperatures soon after the decoupling, that is $T \sim T_D$, which, on account of the equations of motion from the action (\ref{action}), implies a scaling ${\dot b}^{\rm KR}  \sim a(t)^{-3} \sim T^3$, for $T < T_D$, with $a(t) \sim T^{-1}$ the scale factor of the Universe during radiation dominance. Matching with the values (\ref{values}), we then obtain that today the KR axion background is $b_0 = {\mathcal O}(10^{-44})$ meV, implying the the model is consistent with the current constraints of the $b_\mu$ SME parameter today~\cite{smebounds}: $b_0 \, < \, 0.02 \, {\rm eV} , \quad b_i < 10^{-31}~{\rm  GeV}$. 
Moreover, as discussed in \cite{dms}, the above cooling law also implies that the KR axion background is  sufficiently small during the  nucleosynthesis  era, so the latter is not affected.

\section{Discussion}

In this talk we presented a string-inspired model for a LV- and CPTV-induced baryon asymmetry through leptogenesis at early eras of the Universe. 
Unlike the standard CPT-conserving models for leptogenesis, our model does not require mixing among RHN in order to yield the necessary amount of CPV for the generation of phenomenologically acceptable matter-antimatter asymmetry in the Universe. A single species suffices\footnote{Nevertheless, we can straightforwardly extend the scenario to two or three right handed neutrinos, which then, through the seesaw mechanism~\cite{seesaw} can give small masses to the light neutrinos of the SM sector.}, provided 
 the temporal component $b_0$ of the dual of the background field strength of the antisymmetric tensor field is fixed appropriately, together with the RHN mass and the Yukawa coupling of the Higgs portal (\emph{cf.} (\ref{values})).
There are many open issues of course associated with the scenario, the most important of all regarding the suppression of the vacuum energy density, which we only hinted to in this talk and in our previous work~\cite{dms}. A detailed microscopic string model, where such a suppression can occur naturally, is still pending. Nonetheless, we believe that the above-presented class of string-inspired models for  baryogenesis through CPTV leptogenesis constitutes an interesting avenue for future research, which has good prospects to lead to fully realistic models.

Before closing the talk I would like to make some brief remarks on the potential r\^ole of the quantum fluctuations of the KR axion field $b^{\rm KR}(x)$ on  inducing themselves a Majorana mass term for the RHN. Such a scenario has been discussed in \cite{mavropilaftsis} and involves a coupling (via kinetic mixing) of the KR axion with ordinary axions, $a(x)$, which in turn couple to the RHN species via chirality-preserving Yukawa coupling terms in the effective action of the form: ${\mathcal L}_{a} = \int d^4 x \sqrt{-g} \, i \, Y_a \, a (x) \, (\overline N_R^c N_R - \overline N_R\, N_R^c )$. The RHN Majorana mass is then generated radiatively via two loop diagrams involving two graviton and one axion exchanges~\cite{mavropilaftsis}. Light (SM sector) neutrino masses (compatible with the observed oscillations) can then be generated via a seesaw mechanism~\cite{seesaw}, in case there are more than one species of RHN, in a way consistent with the SU$_{\rm L}$(2) gauge invariance of the SM sector. The Yukawa term ${\mathcal L}_a$ breaks the standard axion shift symmetry. In microscopic string theory models such terms may be generated by instanton effects, although the detailed microscopic dynamics is still to be explored.

\section*{Acknowledgements} 

This work is partially supported in part by STFC (UK) under the research grant ST/L000326/1.

\end{document}